\documentclass[10pt,=preprint]{aastex63}
\graphicspath{{./}{figures/}}
\usepackage{subfigure}

\begin{document}
\title{Differential Source Count for Gamma Ray Bursts}

\correspondingauthor{Shreya Banerjee}
\email{shreya.banerjee@fau.de}

\author[0000-0002-0786-7307]{Shreya Banerjee}
\affiliation{Inst. for Quantum Gravity, FAU Erlangen-Nuremberg,
Staudtstr. 7, 91058 Erlangen, Germany}

\author{David Eichler}
\affiliation{Department of Physics, Ben-Gurion University, P.O.Box 653, Beer-Sheva 84105 Israel}
\email{Prof. David Eichler sadly passed away during the elaboration of the new version of this manuscript.}

\author{Dafne Guetta}
\affiliation{Department of Physics, ORT-Braude College, Karmiel, Israel}
\email{dafneguetta@braude.ac.il}



\begin{abstract}
Different forms of long gamma-ray bursts (GRBs) Luminosity Functions are considered on the basis of an  explicit physical model. 
The inferred flux distributions are compared with the observed ones from two samples of GRBs, Swift and Fermi GBM.
The best fit parameters of the Luminosity functions are found and the physical interpretations are discussed.
The results are consistent with the observation of a comparable number of flat phase afterglows and monotonic decreasing ones.

\end{abstract}
\keywords{gamma-ray burst: general}



\section{Introduction}

 Gamma Ray Bursts (GRBs) are among the brightest cosmological
explosions in the universe. 
After their discovery (Klebesadel et al. 1973), it was
shown that there are two types of GRBs (Mazets et al. 1981) - short/hard and long/soft ones. This is
well supported both from observational data and from statistical analysis
(for a survey and the relevant references see, e.g., Meszaros 2006). 

This classification is supported also theoretically as the progentitors of these two GRB classes
are different. Short/hard bursts are given by the neutron star - neutron star mergers, where one or even two neutron stars can be substituted by black holes, and
are observed as macronovae (Tanvir et al. 2013). These sources can also emit gravitational waves (Abbott et al.
2017). The discovery of  a gravitational wave signal from GRB 170817 (LIGO 2017, Goldstein et al 2017, Savchenko et al 2017)
is a confirmation of the hypothesis that short/hard GRBs arose at the final stage of compact binaries.
The long/soft bursts are associated with Ib/c type supernovae (Hjorth et al. 2003). The
idea that the GRBs were coupled to supernovae was formulated in essence simultaneously with the discovery of bursts (Colgate 1968, Colgate 1974).

There are observational hints for further subclasses (Mukherjee et al. 1998, Horvath 1998). From the statistical studies of the dataset of BATSE instrument, a third subgroup of GRBs was identified (Horvath 2002, Hakkila et al. 2003, Hakkila et al. 2004, Horvath et al. 2006), characterized by an intermediate duration. Swift (Veres et al. 2010) and RHESSI (Ripa et al. 2012) confirmed the existence of this third subgroup while 
Fermi did not find any GRBs belonging to this third class (Tarnopolski 2015, Narayana Bhat et al. 2016). Similarly, no
third subgroup is declared to exist both in the Suzaku database (Ohmori et al. 2016),
and in the Konus/WIND catalog (Tsvetkova et al. 2017).
Therefore all hints for more subgroups - beyond the short/hard and
long/soft ones - are controversial and different biases, selection effects should be taken into account (Hakkila et al. 2003, Tarnopolski 2016). 

There are studies claiming the existence of other subgroups - being not identical to the intermediate one, however from the observational point of view there are two sort of GRBs - either
with or without the measured redshift.

In this paper we consider only long/soft GRBs both with and without the measured redshifts. We will also consider a subclass of the long/soft GRBs characterized by low luminosity (LLGRBs).

 An intrinsic luminosity function (ILF) is among the most sought after quantities for any class of astrophysical objects. From quasars to various types of galaxies to supernovae, etc., these distributions provide important insights not only into the physics of the individual objects themselves, but also into the evolution of matter in our universe. 
 
 For GRBs, which are believed to be beamed along a jet axis,  the intrinsic luminosity function also sheds light on the angular emission profile of the GRBs, assuming the observers are distributed isotropically. As only a small, possibly biased fraction of GRBs have observed redshifts at the present time, the derivation of their luminosity function has presented a difficult problem, especially at the low luminosity end. In particular, it is not settled whether low luminosity emission along lines of sight well off the jet axis is dominated by a) material moving at angular offset to the line of sight, or b) by material moving along the line of sight. If these photons would be moving backward, moving more than $1/\Gamma$ from the emitting material in the observer's frame or in the frame of the presumably ultra-relativistic emitting material, this would have enormous implications for models of GRBs, where it is frequently assumed that the observed emission is beamed {\it forward}.

A luminosity function is a measure of the number of objects per unit luminosity and therefore is intimately connected to the energy budget (e.g., mass and rotational energy) and the physical parameters determining the emission mechanism (e.g. density and magnetic field) of the objects. 

Because GRBs are most likely associated with some sort of compact object(s) (e.g., a very massive star MacFadyen et al 2001), or two compact objects, such as neutron stars or black holes, merging (Eichler et al 1989), getting a handle on the GRB luminosity function and event rate could have important consequences for understanding not only GRBs themselves, but other astrophysical problems as well. 

There exists an inevitable scatter in intrinsic luminosities of the GRBs, it is not clear if this scatter is due to relativistic beaming as offset viewers may see diminished luminosity. Scatter in the intrinsic luminosity can cause a selection bias. When the scatter in luminosity is strong, it is important to quantify the effect to distinguish it from true dynamical evolution. Thus, it is important to study luminosity selection effects in GRBs in order to identify the bias. 

Since the paper of Colgate (Colgate 1968), and mainly after the Bepposax detection of afterglows (i.e. after February 28, 1997), there are huge
number of observational
studies about the opening angles and about the jets structures of the
GRBs (i.e. Wang et al 2018).
Today it is sure that the long GRBs are coupled to the supernovae types
Ib and Ic. There exist many theoretical and observational
articles discussing the structures of the jets, supernovae (c.f. Colgate 1968, Meszaros and Rees 2001, Matzner 2003).

However the number of GRBs with known redshift and opening angle is very small in comparison to all the GRBs detected, therefore still it is not possible to have a firm conclusion on the luminosity function form and redshift distribution behaviour of GRBs.

To date, several authors, such as in (Guetta and Piran 2005, Guetta and Piran 2007, Guetta et al 2005, Guo et al 2020, Firmani et al 2005, Tsvetkova et al 2017, Petrosian et al 2015, Pescalli et al 2016, Guetta and Piran 2006), have tried to constrain the luminosity function from the observed fluxes of GRBs. 
Most of these papers consider a broken power law luminosity function,
$N(L)\propto (L/L_0)^{-\alpha}$ for $L<L_0$ and, $N(L)\propto (L/L_0)^{-\beta}$ for $L>L_0$,  and give constraints on the low and high power law index and on the luminosity break. 

In this paper, we consider a physical model for the GRBs and attempt to constrain the implied ILF $N(L)$, using the observed $N(F)$, i.e. the numbers of GRBs  observed per unit flux interval. 
Assuming that GRBs follow the star formation rate (Lloyd-Ronning et al 2002, Yonetoku et al 2004), for each $N(L)$, it is possible to find the implied $N(F)$ and compare it to the observed data.
We also consider all the GRBs with given redshift, estimate the luminosities of these GRBs and compare these data with the GRBs of the entire sample. 

The paper is organized as follows: in Section 2 we describe our sample consisting of the observed GRBs, in Section 3 we describe our physical model and the implied luminosity function forms, in Section 4 we describe the method used to constrain the luminosity function parameters, in Section 5 we present our statistical analysis of the models, in Section 6 we show the results, in Section 7 we compare the results for GRBs with given redshifts with the ones obtained for the entire sample, finally in Section 8 we give our conclusions.

\section{Samples}
In this section we give a detailed description of the long GRBs data used in this paper in order to perform our analysis. One of our samples consists of $918$ long GRBs detected by Fermi Gamma Ray Bursts monitor. The list of the long GRB peak fluxes has been obtained from (Gruber et al 2014, Kienlin et al 2014, Bhat et al 2016, Meszaros and Meszaros 1996, https://heasarc.gsfc.nasa.gov/W3Browse/fermi/fermigbrst.html, https://www.mpe.mpg.de/~jcg/grb.html, https://sites.astro.caltech.edu/grbox/grbox.php). In order to represent graphically  the data of this sample, we have plotted the histogram representing the number of events per unit flux interval (Fig.1).

Another sample used in this paper is the one containing the Swift data obtained from Swift catalogue (https://swift.gsfc.nasa.gov/) and (https://www.mpe.mpg.de/~jcg/grb.html, https://sites.astro.caltech.edu/grbox/grbox.php).
It consists of $365$ long GRBs. Also in this case we have plotted the histogram representing the number of events per unit flux interval (Fig.2).


We have also considered the sample consisting of GRBs with known redshifts. For this, we have considered all the GRBs that have measured redshift from the above samples. The list has been obtained from (Gruber et al 2011, Gruber 2012, https://www.mpe.mpg.de/~jcg/grb.html, https://sites.astro.caltech.edu/grbox/grbox.php) for Fermi GBM. The Fermi GBM sample consists of 33 long GRBs detected by Fermi GBM with know redshifts. For the Swift case, the sample consists of 198 long GRBs obtained from (https://swift.gsfc.nasa.gov/, https://www.mpe.mpg.de/~jcg/grb.html,  https://sites.astro.caltech.edu/grbox/grbox.php). For both the cases, we obtain the histogram for the differential source counts using their peak flux distributions and compared it to the peak flux of the all sample (Fig3).



\section{Description of the physical model and methodology}
The structure and opening angle distribution of GRB jets have impact on their prompt and afterglow emission properties. The jet of GRBs could be uniform, with constant energy per unit solid angle
within the jet aperture, or it could instead be structured, namely with energy and velocity that
depend on the angular distance from the axis of the jet. Thanks to the afterglow detection it was possible to determine the redshift and the opening angle for a small sample of GRBs.
Using the observed data, many theoretical and observational articles discuss the possibility that jets have a structural form that may be connected to the structure of the supernovae, (c.f. Colgate 1968, Meszaros and Rees 2001, Matzner 2003). However the number of GRBs with known redshift and opening angle is very small in comparison to the all sample of the GRBs therefore in this work we try to get some insight about the
still unknown structure of GRBs by studying their luminosity function (Pescalli et al 2015).
In this paper we consider the possibility that the jet is uniform and study the observational implication of this assumption on the luminosity function form. We then compare our results with the ones obtained for the sample of GRBs with known redshift.

We assume that all the jets are identical, i.e. with the same Lorentz factor $\Gamma$ and opening angle $\theta_0$. We also assume that the distribution of observers is isotropic. They would nevertheless be distributed in intrinsic isotropic luminosities $L$ because different observers would see them at different viewing angles. 
We distinguish three possible physical scenarios:

\textbf{ Optically Thick Scenario:}  The head on observers do not see any emission as the material in the jet trap the photons to exit the jet region, therefore the jet is optically thick. Observers outside the jet will see different emission according to their position relative to the jet opening angle. In (Banerjee et al 2020), paper1, the luminosity function was calculated numerically as a function of the jet opening angle and the Lorentz factor. In order to test the models in paper1 we employed a statistical test based on median luminosity. We have shown there and in other papers (Eichler 2017) that for this physical case, the ILF can be approximated as function $N(L)dL \propto L^{-\alpha}dL$ where $\alpha=1.33$ for  observers slightly off the edge of the jet and $\alpha=1.25$ for observers far from the edge. 

\textbf{ Optically Thin Scenario:} For head-on observers, the luminosity function for a set of identical GRBs would essentially be a delta function $\delta(L-L_{max})$ if the beam is optically thin. This case is excluded from observations because it predicts more bright bursts than what is observed. However we may have observers outside the optically thin jet. In this case the ILF can be approximated by the sum of a delta function (observers head on the jet) and as function $N(L)dL \propto L^{-\alpha}dL$ where $\alpha=1.33$ for  observers slightly off the edge of the jet and $\alpha=1.25$ for observers far from the edge.

\textbf{Partially Optically Thick Scenario:} 
When the shell is optically thick, observers outside the jet see the same LF as the observers outside an optically thin shell because they are seeing photons that are scattered backwards in the frame of the shell. However observers within the opening angle of the jet cannot see photons backscattered from material moving directly at them. Therefore there is a "blind spot". These observers see backscattered photons only from jet material outside the blind spot. The calculation of the blind spot size, $\theta_{\rm BS}$, is outlined in  (Eichler 2014). It shows that the minimum angle of the blind spot depends on the optical depth of the shell and on the distribution of circumburst material. 
The blind spot means that there is a minimum offset between the observed material and the line of sight. This is in contrast to observers outside the jet who can come arbitrarily close to the edge of the jet. If the illumination is in pulses, the observers within the jet see a light echo from those parts of the jet that are outside the blind spot. Most of the time integrated signal comes from an annulus just outside the blind spot and comparable in solid angle to the solid angle of the blind spot itself. If the observer is well within the jet, the annulus completely surrounds the blind spot, but if the blind spot touches the edge of the jet then the annulus is only partial and the observer very close to the edge of the jet sees an annulus that it is only about half of the circle. The geometry of this model is discussed in detail in (Vyas, Peers and Eichler 2021). 
For observers just outside the jet, the luminosity function can be approximated by a function $N(L)dL \propto L^{-\alpha}dL$ where $\alpha=1.33$ for  observers slightly off the edge of the jet and $\alpha=1.25$ for observers far from the edge as described above.

For observers inside the jet the luminosity function depends on the distribution of blind spot sizes.

\textbf{Scale free distribution of blind spot sizes:}
If the distribution is scale free then it is reasonable to represent this by another power law (if the blind spot size is bigger than the jet opening angle then it become meaningless because the emission the observers see is identically zero). The total luminosity function is then the sum of the luminosity functions for observers outside the jet and the ones inside the jet. So we always write the ILF as the sum of two terms, $N(L)=N_{\rm in}(L)+N_{\rm out}(L)$. The $N_{\rm out}(L)\sim ((L/L_{\rm max})^{-\alpha}$ as discussed above. If the distribution of the a blind spot sizes is scale free, then it is reasonable to approximate it as  $N(L)_{\rm in}\sim (L/L_{\rm max})^{-\beta} $. In this case, the total ILF can be represented by  $N(L)\propto ((L/L_{\rm max})^{-\alpha}+\epsilon (L/L_{\rm max})^{-\beta} )$.
The total ILF can be also approximately represented by a broken power exempt for a region where both components give comparable contribution. For this reason our results that come from physical model resemble previous works that consider a broken power law as the best form for the GRBs luminosity function. 

\textbf{Characteristic scale distribution of blind spot sizes:}
If the blind spot has a characteristic scale (the opposite extreme is scale free) i.e. $\theta_{\rm BS}\sim 3/\Gamma$ as estimated in (Eichler 2014), then the $N_{\rm in}(L)$ should be dominated by a peak at the corresponding luminosity, $L_*$. In this case, the total ILF can be represented by the sum of a power law and a delta function $N(L)\propto \delta(1-L/L_*)$. 

In this paper we consider the possible ILF forms described above and test these models against observed {\it flux} distributions. In order to constraint our model parameters we use the Fermi GBM and Swift sample of GRBs described in section 2, thereby finding the physical model that explains the data most accurately.

\section{Differential Count}
Let $N(L,z)$ be the luminosity distribution function(LDF) of a class of emitters: the rate of detected GRBs per unit comoving volume per unit luminosity at a redshift $z$. For a GRB with luminosity $L$ (in units photons/s) being at redshift $z$, 
if there are
given two photon-energies E1 and E2, where $E1 < E2$, then the
flux $F$ (in units photons/(cm$^2$s)) of the photons with energies
$E1 \lesssim E \lesssim E2$ detected from a GRB having a redshift $z$ is given by (Meszaros et al 2011)

\begin{equation}
F = \frac{(1+z)^{(2-\gamma)}L}{4\pi D_L(z)^2},
\end{equation}
where $D_L(z)$ is the luminosity distance. Also, if $F$ and $L$ are known, then the redshift $z$ is calculable from $D_L(z) = \sqrt{L(1+z)^{2-\gamma}/(4\pi F)}$. 
The parameter $\gamma$ is the spectral index of the peak flux. 
 From (Yu et al 2016, Nava et al 2018, Gruber et al 2014, Kienlin et al 2020) and GBM and Swift Burst catalogue we take the average spectral index, $\gamma \sim 1.6$, of the peak fluxes considered in our sample. We have checked that our results do not change significantly by changing $1<\gamma <2$.

For a fixed $L$, $N(>F)$
is the number of emmitters being at smaller redshifts than $z$ obtained from
$D_L(z) = \sqrt{(1+z)^{(2-\gamma)}L/(4\pi F)}$. We denote this redshift as $z(L, F)$. Then the all
observed emitters $N(>F)$ will be given by the double integral

\begin{equation}
N(>F)= \int_{L_{min}}^{L_{max}} \int_{0}^{z(L,F)}N(L, z)\frac{dV(z)}{dz}\frac{1}{1+z}dzdL
\label{1}
\end{equation}
where $dV/dz$ is the comoving volume element per unit of redshift, the
factor $(1 + z)^{-1}$ accounts for the cosmological time dilation.



In our analysis we consider the evolution of the luminosity function with redshift such that $N(L,z)=N(L)N(z)$ (as in (Lloyd-Ronning et al 2002, Yonetoku et al 2004)), where the luminosity function $N(L)$ represents the relative fraction of bursts with a certain luminosity and the GRB formation rate $N(z)$ which represents the number of bursts per unit comoving volume and time as a function of redshift.
Assuming that the life-time of the
GRBs progenitor massive stars are small enough, we can consider the possibility 
that $N(z)$ at redshift $z$ follows the star formation rate (Meszaros et al 2006).

 However it is important to mention that there are evidences that the evolution of  GRBs may deviate from the star formation rate $\rho(z)$ (Lloyd-Ronning et al 2002, Petrosian et al 2015)
i.e.,
\begin{equation}\label{Rcj}
N(z)=\rho_o\rho(z),
\end{equation}
where the star formation rate is (Madau 2014)
\begin{equation}
\rho(z)=\frac{(1+z)^{2.7}}{1+[(1+z)/2.9]^{5.6}},
\end{equation}
 Here $\rho_0 $ is the local rate of GRBs. 

Following the description of the physical model given in Section 3 we consider different choices of the LF, $N(L)$, and compare the the predicted $N(>F)$ with the observed data as described in Section 2. 

\section{Statistical Analysis}

In this section we test our model described in Section 3 against observed data (described in Section 2) for different scenarios.

\textbf{Optically Thick Scenario.}

As our first choice we consider the case where the GRBs are observed out of the jet (they can be close by the jet or far away). In this case the luminosity function of GRB peak luminosities $L$, defined as the comoving space density of GRBs in
the interval $L$ to $L + dL$, is
\begin{equation}
{\cal N}(L)=c_0 \Big( \frac{L}{L_{max}}\Big)^{-\alpha}
\end{equation}
where $c_0$ is a normalization constant with dimension of $1/L$ so that the integral over the luminosity function equals unity.

Considering the evolution of the luminosity function with redshift such that $N(L,z)=N(L)N(z)$ (as in (Lloyd-Ronning et al 2002, Yonetoku et al 2004)), we get 
\begin{equation}
N(L,z)=\rho_0c_0\Big( \frac{L}{L_{max}}\Big)^{-\alpha}\frac{(1+z)^{2.7}}{1+[(1+z)/2.9]^{5.6}}
=N_0\Big( \frac{L}{L_{max}}\Big)^{-\alpha}\frac{(1+z)^{2.7}}{1+[(1+z)/2.9]^{5.6}}
\label{N}
\end{equation}

Here $\alpha$ is the spectral index with $1<\alpha<2$. The condition $1 < \alpha < 2$ implies that while most of the power is in the brightest sources, there are more dim sources than bright ones. 

In order to fix the normalization constant, $N_0$ is fixed by the value of the local GRB rate as explained earlier., we integrate Eq. \ref{N} over $L$ and $z$.  We get the rate of GRBs as $2$ per day for $\rho_0\approx 5.5 Gpc^{-3}yr^{-1}$, consistent with that observed by Swift/GBM (Pescalli et al 2015).

\textbf{Optically Thin Scenario.}

We consider the case where observers are outside the optically thin jet. In this case the $N(L,z)$ can be approximated by 

\begin{equation}
 N(L,z)=N_0 \Big(\Big(\frac{L}{L_{max}}\Big)^{-\alpha}+\delta(1-\frac{L}{L_{max}}\Big)\Big)\frac{(1+z)^{2.7}}{1+[(1+z)/2.9]^{5.6}}
 \label{case3}
\end{equation}
The delta function represents the contribution of jets of finite solid angle observed head on.

\textbf{Partially Optically Thick Scenario.}
As described in detail in section 3 when the shell is partially optically thick the total luminosity function is then the sum of the luminosity functions for observers outside the jet and the ones inside the jet. So we always write the ILF as the sum of two terms, $N(L)=N_{\rm in}(L)+N_{\rm out}(L)$. 
 We distinguish two possible extreme cases where the distribution of the blind spot sizes is scale free or has a characteristic size. 
 
\textbf{Scale free distribution of blind spot sizes.}

In this case, the total ILF can be approximated by  $N(L)\propto ((L/L_{\rm max})^{-\alpha}+\epsilon (L/L_{\rm max})^{-\beta} )$.

In this case $N(L,z)$ can be written as:

\begin{equation}
N(L,z)=N_0\Big(\Big(\frac{L}{L_{\max}}\Big)^{-\alpha}+\epsilon \Big(\frac{L}{L_{max}}\Big)^{-\beta} \Big)\frac{(1+z)^{2.7}}{1+[(1+z)/2.9]^{5.6}}
\label{6.2}
\end{equation}
Their relative contribution is determined by the dimensionless constant $\epsilon$. $N_0$ is the normalization constant fixed in a way similar to earlier cases.  Both $\alpha$ and $\beta$ have their values between $1$ and $2$.
In this case the total ILF can be also approximately represented by a broken power with parameters $\alpha$ and $\beta$ and a luminosity break at $L_0$  exempt for a region where both components give comparable contribution.


We consider different values for $\alpha,\ \beta,\ L_0$ and looked for the combination that fits the data most accurately.

\textbf{Characteristic scale distribution of blind spot sizes.}

If the blind spot has a characteristic scale, then the $N_{\rm in}(L)$ should be dominated by a peak at the corresponding luminosity, $L_*$. In this case, the ILF is a sum of a power law and a delta function around  $L_*$. 
The $N(L,z)$ luminosity function is now given by
\begin{equation}
 N(L,z)=N_0 \Big(\Big(\frac{L}{L_{max}}\Big)^{-\alpha}+\delta\Big(1-\frac{L}{L_{*}}\Big)\Big)\frac{(1+z)^{2.7}}{1+[(1+z)/2.9]^{5.6}}
 \label{case3}
\end{equation}
The delta function represents the contribution of jets of finite solid angle observed head on.

\section{Results for the all GRB sample}

We compare the theoretical models with the observed data by 
plotting the $logN-logF$ obtained from the different ILFs considered in the previous section versus the observed $logN-logF$. Our results are shown in Fig.1 (for the GBM sample) and Fig.2 (for the Swift sample). 
In order to have a quantitative result on the goodness of the fit, we use the method of $\chi^2$ minimization and p value. The level of significance is a predefined threshold, which in our case has been set as $0.05$.  The results, using both Fermi GBM and Swift data, are reported in Table \ref{pvalue1} and \ref{pvalue2} respectively.

\textbf{Optically Thick Case}

\begin{figure}[h!]
\centering
\includegraphics[width=12cm,height=10cm]{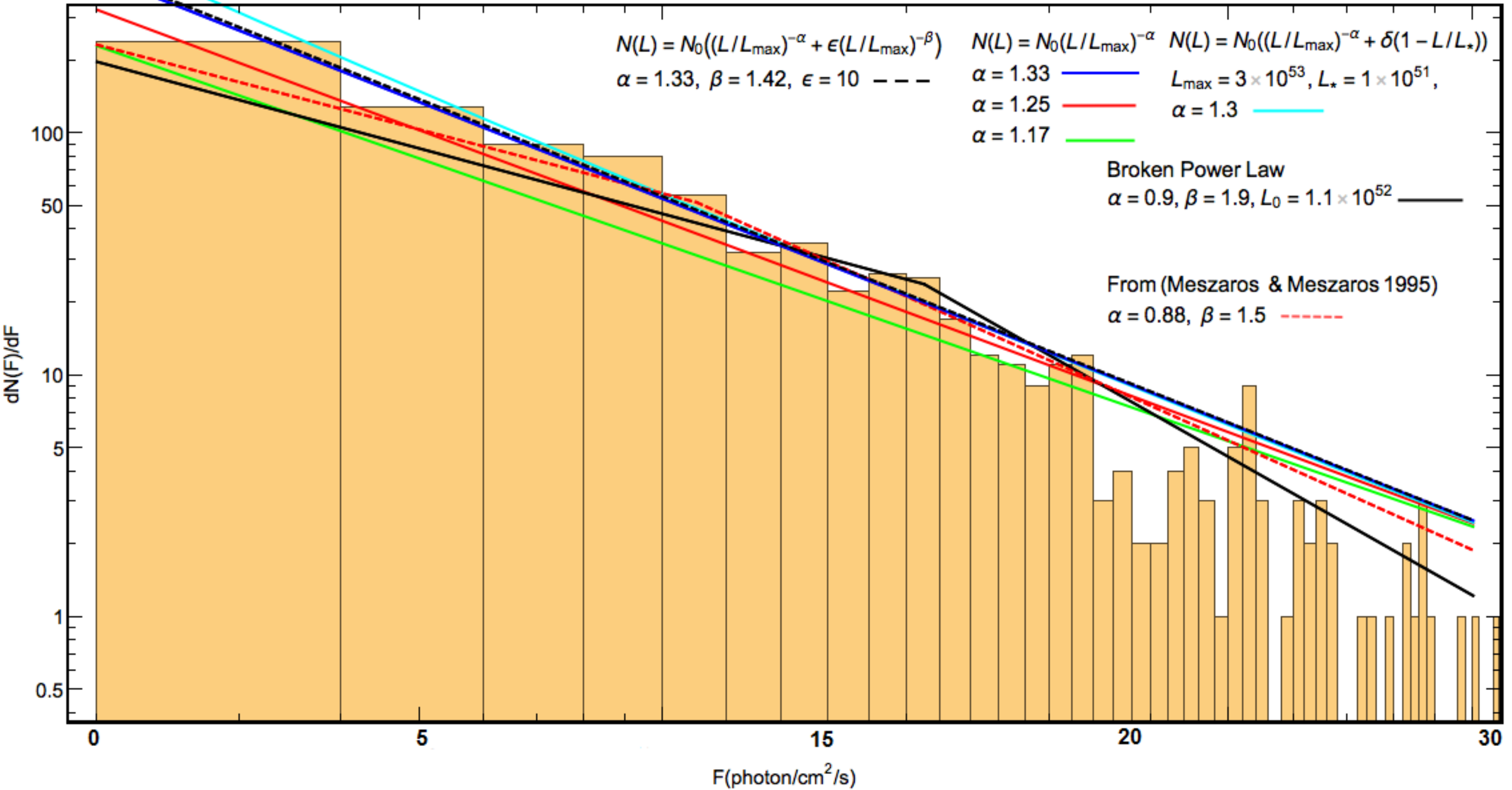}
\caption{Variation of ${\cal N}$ with $F(photon/cm^2/s)$ for the models considered in the paper. Optically thick case: Blue curve-$\alpha=1.33$, red curve-$\alpha=1.25$, green curve-$\alpha=1.17$. 
Partially optically thick case where with a scale free distribution of blind spot sizes-($\alpha=1.33,\ \beta=1.42,\ \epsilon=10$)-black dashed curve.
Partially optically thick case where with characteristic scale distribution of blind spot sizes 
represented by a delta function plus single power law-($\alpha=1.3,\ L_{max}=3\times 10^{53},\ L_{\star}=1\times 10^{51}$)-cyan curve.
For comparison with other works in the literature we have also approximated the ILF as
a broken power law-($\alpha=0.9,\ \beta=1.9,\ L_0=1.1\times 10^{52}$)-solid black curve. Fig 1 from (Meszaros and Meszaros 1996)-($\alpha=0.88,\ \beta=1.8$,\ Break peak flux $F_b=10$photon/cm$^{-2}$/s-red dashed curve. GBM data are given by the histogram}
\label{graph2}
\end{figure}

\begin{figure}[h!]
\centering
\includegraphics[width=12cm,height=10cm]{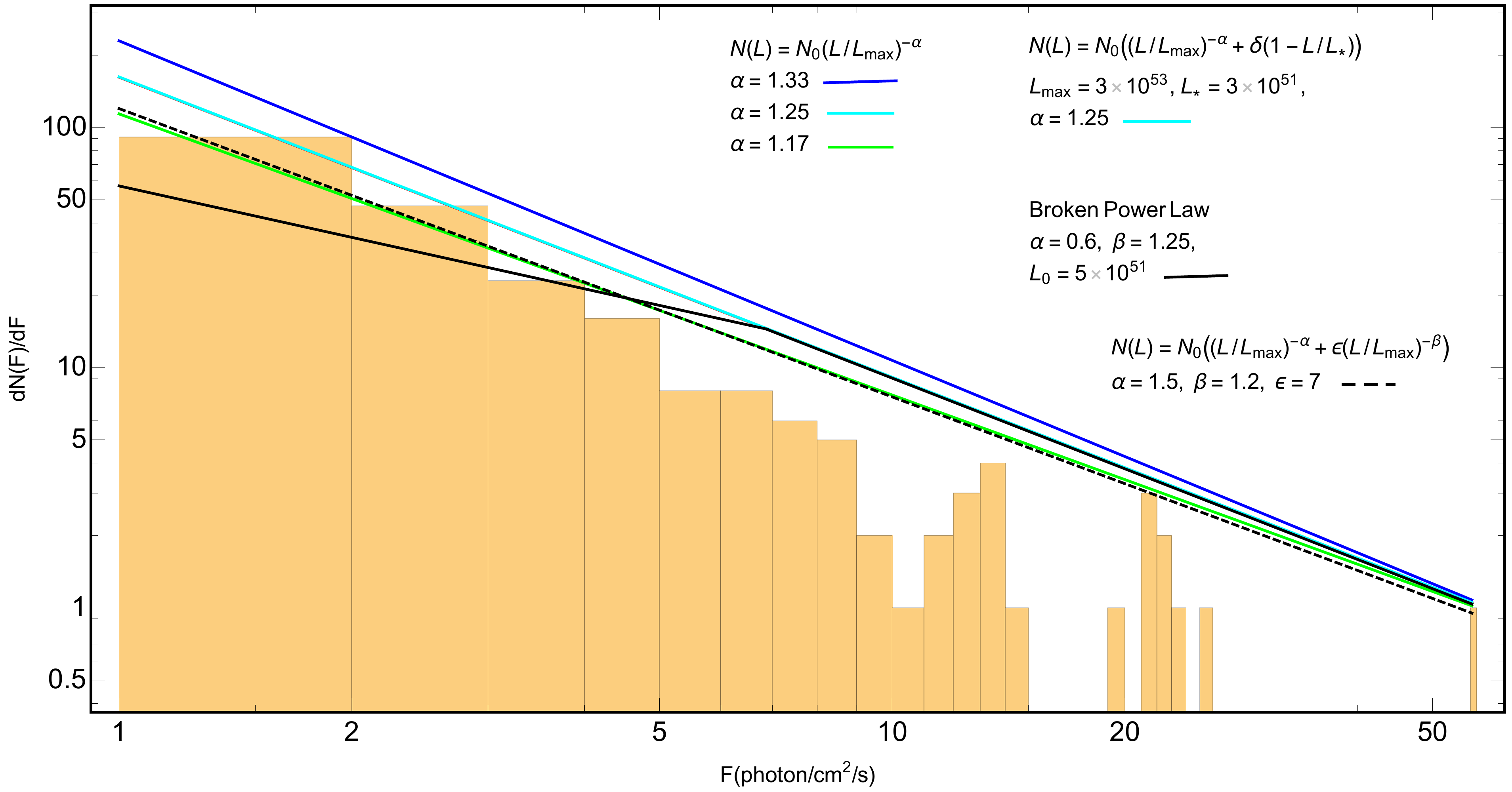}
\caption{Variation of ${\cal N}$ with $F$ for the models considered in the paper. Optically thick case: Blue curve-$\alpha=1.33$, red curve-$\alpha=1.25$, green curve-$\alpha=1.17$.
Partially optically thick case where with a scale free distribution of blind spot sizes-($\alpha=1.5,\ \beta=1.2,\ \epsilon=7$)-black dashed curve.
Partially optically thick case where with characteristic scale distribution of blind spot sizes 
represented by a delta function plus single power law-($\alpha=1.25,\ L_{max}=3\times 10^{53},\ L_{\star}=3\times 10^{51}$)-cyan curve.
For comparison with other works in the literature we have also approximated the ILF as
a broken power law-($\alpha=0.6,\ \beta=1.25,\ L_0=5\times 10^{51}$)-black curve. Swift data are given by the histogram.}
\label{graph3}
\end{figure}

In Figs. \ref{graph2},\ref{graph3}, we have plotted the variation of ${\cal N}$ with $F$ for $\alpha=1.33,\ 1.25,\ 1.17$. 
As we can see from Figure 1,2 and Table 1,2, the single power law model does not fit the data (both for the GBM and Swift sample), which is clearly seen from the low p values. This result was expected as a single power law LF cannot represent the entire sample of the GRBs. As already noted in (Meszaros \& Meszaros 1995, 1996)  it holds that for the largest fluxes the slope  is around -$1.5$ (both for the standard candle assumption and for the power law assumption), but for
smaller fluxes there is a line with slope around $-1$ or so on the $logF-logN$
graph. Therefore the optically thick case can be excluded from our analysis.

\textbf{Partially Optically Thick Case: Scale Free}

For the partially optically thick case with a scale free in the distribution of the blind spot sizes using the observed Fermi GBM data, we find the best fit values of the parameters as $\alpha=1.33_{-0.02}^{+0.02},\ \beta=1.42_{-0.04}^{+0.04},\ \epsilon=10_{-5}^{+5}$.
Similarly, using Swift data, we find the best fit values of the parameters as $\alpha=1.5_{-0.2}^{+0.02},\ \beta=1.2_{-0.04}^{+0.04},\ \epsilon=7_{-5}^{+5}$. 
As mentioned above the partially optically thick case may be approximated as a broken power law and the best fit values are $\alpha=0.9,\ \beta=1.9,\ L_0=1.1\times 10^{52}$ for Fermi GBM. For comparison, we have also plotted the curve in Fig. 1 from (Meszaros and Meszaros 1996) that was obtained for the values $\alpha=0.88,\ \beta=1.5,\ F_b=10 ph/cm^{2}/s$ ($F_b$ is the value of peak flux at the break obtained from the reference, this corresponds to the luminosity break in the ILF distribution). As we can see from Table 1 this set of values for the parameter of the broken power law ILF is also good for the GBM data.
As we can see from Fig.2 the Swift GRBs are characterized by having lower fluxes than the GBM ones, therefore the ILFs that fit the GBM sample do not fit the Swift sample.
We find that the best fit values for the Swift data are $\alpha=0.6,\ \beta=1.25,\ L_0=5\times 10^{51}$ for broken power law.
As we can see from our results the partially thick model with a free scale in the blind spot angle is the one that best fit the data.

An interesting information is the ratio between the observers outside the jet and the ones inside. This can be estimated by convolving the $N(L)_{\rm in}$ and $N(L)_{\rm out}$, with observed volume as explained in Eq. 3. For the best fit values given above this ratio is $1.2$. Thus we see that with these parameter values, the observed GRBs are almost equally distributed outside and inside the jet. In order to check the sensitivity of our results to the $\epsilon$ value we have plotted the change in p value as we vary $\epsilon$ for both Fermi GBM and Swift data (Fig. 4). As we can see, for the best fit values of $\alpha,\ \beta$, the best p values are for a range of $\epsilon$ between $5-15$. The ratio of observers outside the jet and inside the jet is $\sim 1-5$ respectively, implying that a similar number of observers outside and inside the jet is expected.

\textbf{Partially Optically Thick Case: Characteristic scale}

In Figs. \ref{graph2},\ref{graph3}, we have shown the curve that best fit the Fermi GBM and Swift data respectively for a delta function around the best fit values of the break luminosity $L_{\star}$ and $\alpha$. For the present case, the best fit curve is obtained for $\alpha=1.33,\ L_{max}=3\times 10^{53},\ L_{0}=1_{-5}^{+1}\times 10^{51}$. The corresponding $\chi^2$ and p values have been shown in Table \ref{pvalue1}.
As we can see from the figures and the tables the fit in this case is really bad, therefore this case can be excluded.

\section{GRBs with given redshift}

The increasing number of bursts with measured redshift allows to 
estimate the luminosity function of GRBs with increasing
confidence. In order to check our results we have considered the GRB sample as described in Section 2 for known redshift. We have compared their peak flux distribution to the one of the all GRB sample. 

\begin{figure}[ht]
\begin{subfigure}
  \centering
  \includegraphics[width=9cm,height=10cm]{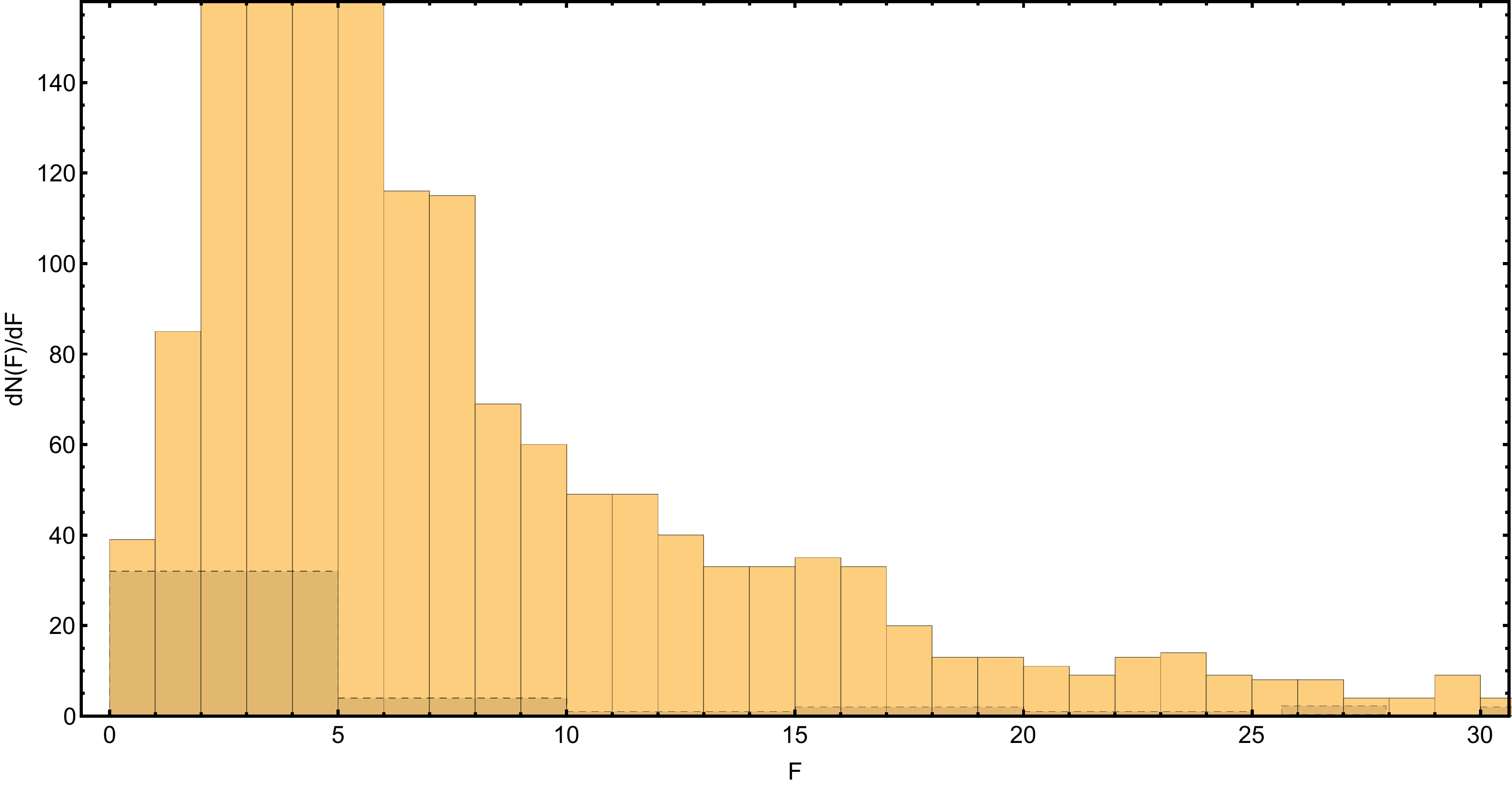}  
\end{subfigure}
\begin{subfigure}
  \centering
  \includegraphics[width=9cm,height=10cm]{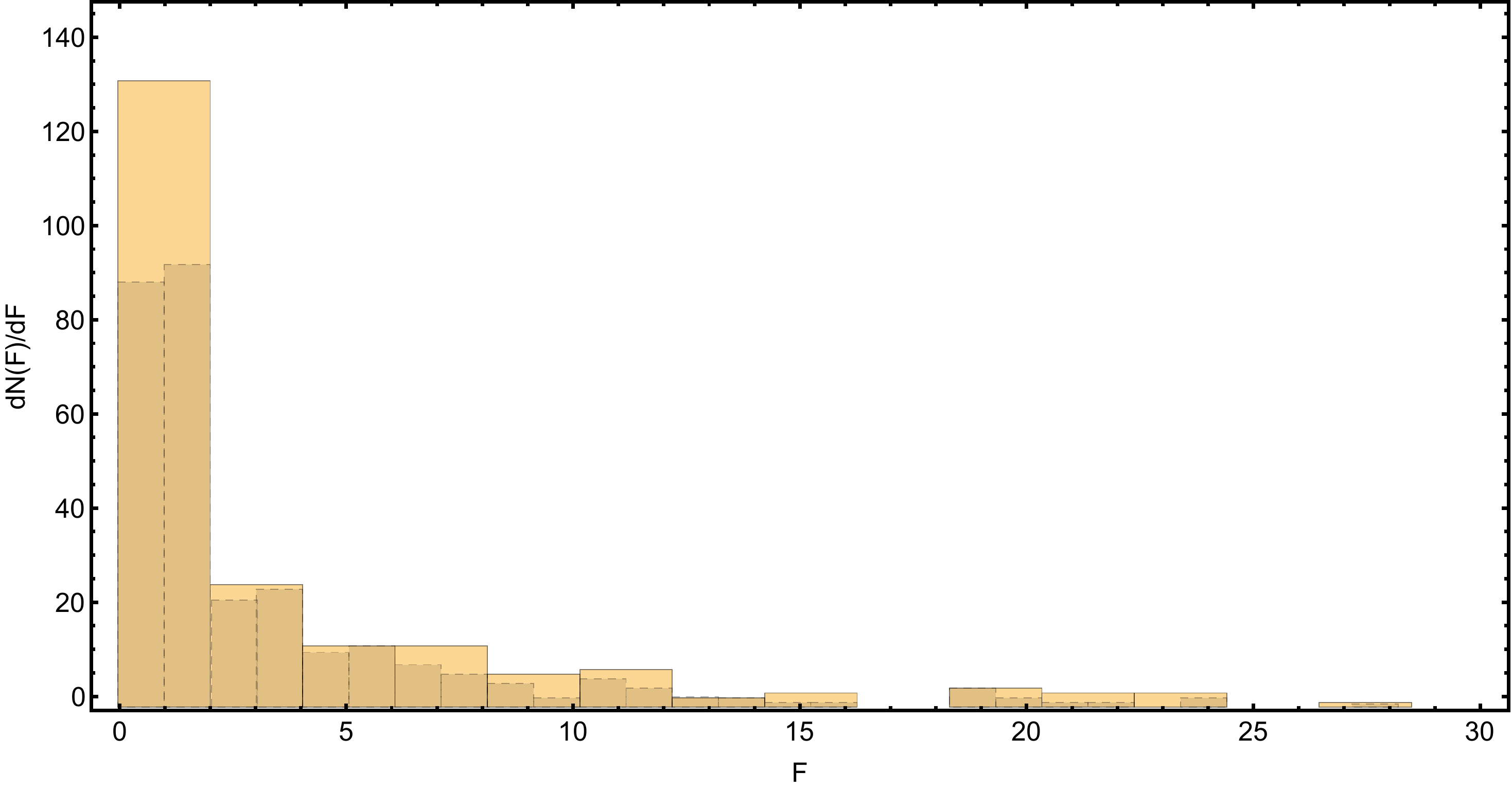}  
\end{subfigure}
\caption{(a)GRB Fermi GBM data- complete set-yellow box, GRB with known redshift-dashed box. (b) GRB Swift data-complete set-yellow box, GRB with known redshift-dashed box. }
\label{graph4}
\end{figure}

The results are reported in Fig. \ref{graph4}, where we have plotted the differential source count for the entire sample (yellow box) and for the sample with known redshifts (grey dashed box), for comparison. The sample of GBM bursts with known redshift is a small fraction of the all GBM sample. As we can see from Fig3 (left box) the GBM bursts with known redshift have typically much smaller fluxes than the GRBs of the all sample.
In the Swift case the percentage of GRBs with known redshift is very high and the fluxes of GRBs without redshift and the ones with redshifts are comparable.

In order to quantify this difference we have fit the GBM and Swift data for GRBs with know redshifts using the broken power law model. We find for the GBM data, the values of the broken power law parameters that fit the data most closely are $\alpha=1.5,\ \beta= 0.5,\ L_0=3\times 10^{51}$ as shown in the last row of Table 1. As we can see, while a low value of $\alpha$ and high value of $\beta$ fits the GBM data for whole GRB sample, the result is opposite if we consider the sample consisting of only known redshifts, giving a fit not as good as when we consider the complete sample.
The results for the Swift case have been shown in the last row of Table 2. For this case, however, we see that the best fit values are nearly same ($\alpha=0.7,\ \beta=1.2$) as for the entire sample. 

Comparing the all Swift sample with the all GBM sample, we see that in the Swift sample there are more low-fluxes GRBs than in the GBM, therefore the luminosity functions that best fit the GBM data do not fit well the Swift data (see Table 1 and Table 2). 
The difference in the two samples and between GRBs with measured redshift vs GRBs without measured redshift may be due to selection or instrumental effects.
However it is possible that the low flux GRBs (LLGRBs) may represent a different population than the standard GRBs (e.g. Virgili et al 2009, Daigne and Mochkovitch 2007, Guetta and Della Valle 2007, Levan et al 2014 (Fig.2)). 

At lower luminosities the detection rate drops to only a few events which are, however, representative of a large local density of GRBs (e.g. Soderberg et al 2006, Pian et al 2006). There are observational evidences that several of these LLGRBs are coupled to SN-Ib/c (Guetta and Della Valle 2007).
The assumption made in our model, that all GRBs have the same jet angle, may be relaxed for the LLGRBs. Indeed there is the possibility
that the low luminous bursts have wide opening angles to account for their low
luminosity (Guetta and Della Valle 2007, Pescalli et al 2015).
These GRBs may also have a different jet structure than the standard GRBs that may be connected to the structure of the supernova itself  (c.f. Colgate 1968, Meszaros and Rees 2001, Matzner 2003). A possible model for the LLGRBs population can be the "chocked" model (Senno et al 2016). Even if accretion driven engines were present in a significant fraction of SNe, they would be choked by the progenitor envelope and either no or a rather weak GRB might be seen. Such a scenario has been suggested as a plausible explanation for the low-luminosity GRBs (Bromberg et al 2011), which, while weak in $\gamma$ -rays,
often host SNe very similar to those seen in the prototypical SN/
GRB 030329 (e.g., Hjorth et al 2003, Pian et al 2006).

The analysis performed in our paper concentrate on the standard long GRBs as our models can fit in a reasonable way the standard long GRBs peak flux distribution.

\begin{table*}[t]
\centering
\begin{tabular}{ |c|c|c| } 
\hline
\quad \quad \quad \quad $\alpha$  \quad \quad \quad \quad  &  \quad \quad \quad \quad $\chi^2$  \quad \quad \quad \quad & \quad \quad \quad \quad p value \quad \quad \quad \quad\\ 
\hline
$1.17$ & $ 5.5$ 
 & $0.01$ \\ 
\hline
 $1.25$ & $6.7$ 
  & $0.009$ 
  \\
\hline
$1.33$ & $8.4$ 
 & $.004$ 
 \\ 
\hline
 $\alpha=0.9,\ \beta= 1.9,\ L_0=1.1\times 10^{52}$ & $1.36 
 $  & $0.24 
 $ \\
 \hline
  $\alpha=0.88$,\ $\beta=1.5$,\ $F_b=10\ photon-cm^{-2}s^{-1}$ & $1.52$  & $0.21$ \\
 \hline
  $\alpha=1.33,\ L_{max}=3\times 10^{53},\ L_{\star}=1_{-5}^{+1}\times 10^{51}$ & $9.1 
  $  & $0.002 
  $ \\
  \hline
  $\alpha=1.33_{-0.02}^{+0.02},\ \beta=1.42_{-0.04}^{+0.04},\ \epsilon=10_{-1}^{+2}$ & $8.8 
  $  & $0.003 
  $ \\
 \hline
 $\alpha=1.5,\ \beta= 0.5,\ L_0=3\times 10^{51}$ 
 & $6.1$ & $0.05$\\
 \hline
\end{tabular}
\caption{Table containing the $\chi^2$ and p values for the models considered in the paper obtained comparing the predictions with the {\it Fermi GBM} data shown in Fig. \ref{graph2}. The last row shows the best fit values for {\it Fermi GBM} data for GRBs with known redshifts.}
\label{pvalue1}
\end{table*}
\begin{table*}[t]
\centering
\begin{tabular}{ |c|c|c| } 
\hline
\quad \quad \quad \quad $\alpha  \quad \quad \quad \quad $ &  \quad \quad \quad \quad $\chi^2$  \quad \quad \quad \quad & \quad \quad \quad \quad p-value \quad \quad \quad \quad\\ 
\hline
$1.17$ & $ 5.24 
$ & $0.07 
$  \\ 
\hline
 $1.25$ & $10.34 
 $ & $0.001 
 $  \\
\hline
$1.33$ & $ 11.12 
$ & $0.0008 
$  \\ 
\hline
  $\alpha=0.6,\ \beta=1.25,\ L_0=5\times 10^{51}$ & $1.81
  $ & $0.178
  $ \\
 \hline
  $\alpha=1.25,\ L_{max}=3\times 10^{53},\ L_{\star}=3_{-7}^{+1}\times 10^{51}$ & $10.34 
  $ & $0.001 
  $ \\
  \hline
  $\alpha=1.5_{-0.2}^{+0.02},\ \beta=1.2_{-0.04}^{+0.04},\ \epsilon=7_{-5}^{+5}$ & $6.23 
  $ & $0.04 
  $ \\
 \hline
  $\alpha=0.7,\ \beta=1.2,\ L_0=7\times 10^{51}$ & $1.67
  $ & $0.2
  $\\
  \hline
\end{tabular}
\caption{Table containing the $\chi^2$ and p values the models considered in the paper obtained comparing the predictions with the {\it Swift} data. The last row shows the best fit values for {\it Swift} data for GRBs with known redshifts.}
\label{pvalue2}
\end{table*}
\begin{figure}[h!]
\centering
\includegraphics[width=12cm,height=10cm]{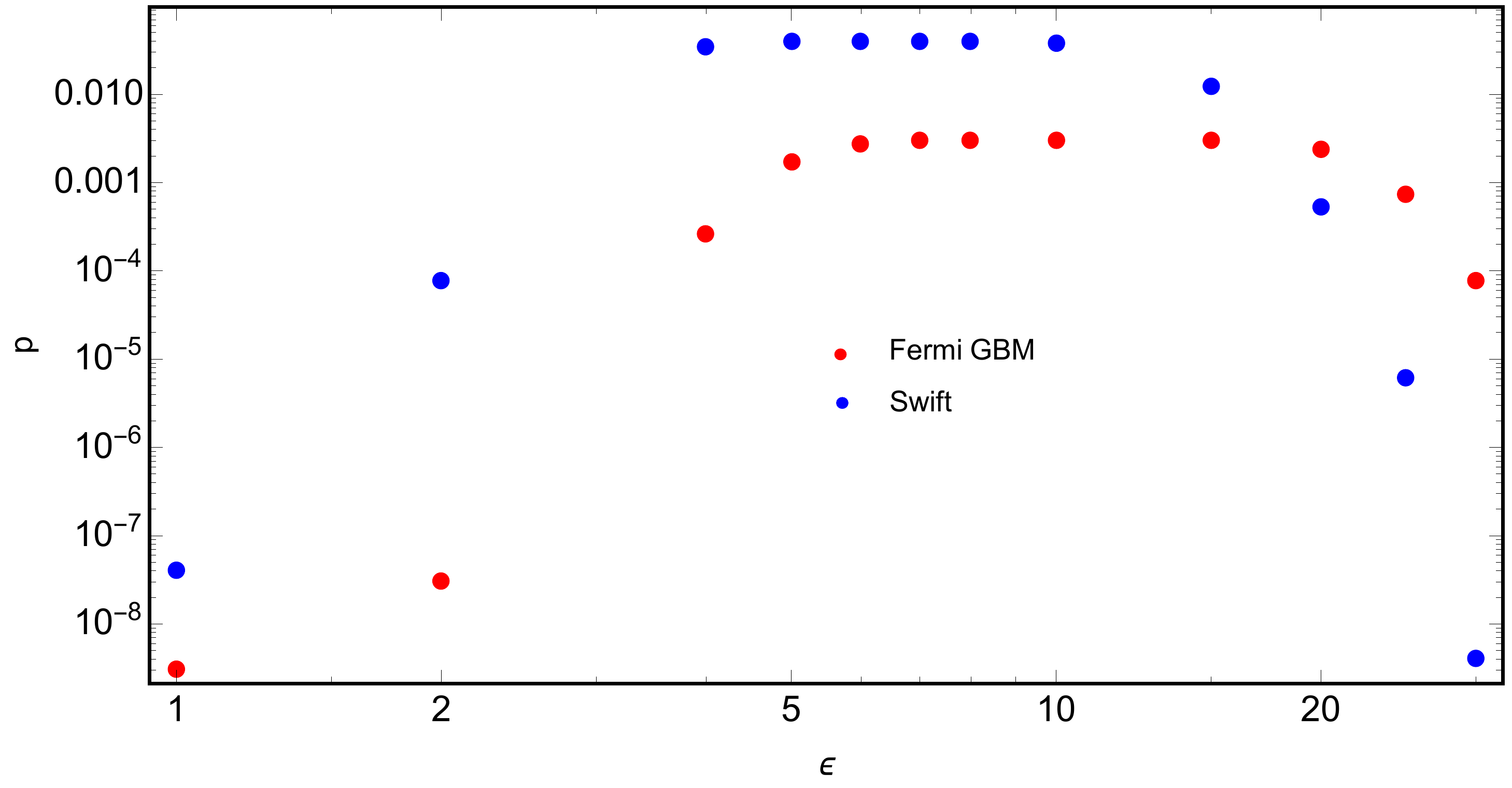}
\caption{Variation of $p$ with $\epsilon$ for sum of two power laws.}
\label{epsilon}
\end{figure}

\section{Conclusion}

In this paper we tried to give a physical interpretation to the intrinsic luminosity function of GRBs.
We have chosen a class of physical models described by Eichler (2014) and references therein. For simplicity we have considered a constant bulk Lorentz factor and an opening angle $\theta_0$.
For each model we studied the behaviour of the differential source count ${\cal N}$ for a given luminosity function  and compared it with the flux distribution of Fermi GBM data and Swift data.
One possible model is a shell with a baryonic material illuminated from behind possibly optically thick. The GRB Intrinsic Luminosity Function consists of a contribution of $N_{\rm out}(L)$ for observers outside the jet and $N_{\rm in}(L)$ inside $\theta_0$. $N_{\rm out}(L)$  can be approximated as $N(L)dL \propto L^{-\alpha}dL$ where $\alpha=4/3$ for observers just outside the jet and $\alpha=5/4$ for observers far from the jet.
We find that a simple power law cannot fit well the data and a break in the luminosity function is needed. Therefore we can exclude the optically thick jet model.

The optically thin case can be represented by a ILF that is a sum of a power law (for the observers out from the jet) and a delta function that peak at $L_{\rm max}$ (for observers that are head on with the jet). As it is clear from Fig. 2 the fit is very bad in this case as the p-value is very low. Therefore the optically thin case can be excluded.

In the case that the jet is partially optically thick we may have a blind spot region inside the jet.
For observers inside the jet, the luminosity function distribution, $N_{\rm in}(L)$, depends on the distribution of blind spot sizes because the optical depth is large when the shell is illuminated from behind. In this case the blind spot hides material within it. In this case the ILF can be represented by the sum of  $N_{\rm in}(L)$ and  $N_{\rm out}(L)$ or by a broken power law. 

Several analysis on the the study of the LF of GRBs have been already performed in the literature and we confirm that the broken power law function is the one that best fit the data. We have tried to connect this LF function to the physical model of a uniform jet with blind spots of different sizes. This is the original part of our work.
As it is difficult to know the distribution of blind spot sizes, we considered both extremes. From our analysis we find that there is no typical size of the blind spot and that a distribution of blind spot angles should be considered.

We find that for the best fit parameters the number of observed bursts from outside the GRB jets is comparable to the number of the GRB jets observed within the GRB angle. This is quite consistent with the observation that the number of GRBs with flat phase afterglow is comparable to the number of GRBs with monotonic decreasing afterglow and the hypothesis that the observers outside the jet see flat phase afterglow while those inside the jet see monotonic decreasing afterglow.

We have considered the GRBs that have measured redshift and checked their distribution. We find that, for the GBM sample, the GRBs with known redshift have lower peak fluxes than the all GRBs in the sample. In the Swift case both GRBs with known and unknown redshift have low peak fluxes. It may be that there is a class of low luminous GRBs that belongs to a different population of GRBs, i.e. with different jet structure and opening angle. Therefore the luminosity function that fit the low peak flux sample may be different than the one found for the standard GRBs.

We find that the best fit LF for the standard GRBs is in good agreement with  values found in the literature. In particular with (Meszaros \& Meszaros 1995). In this sense our work resemble previous works done on the subject, giving an original physical interpretation on the ILF choices.

\section{Acknowledgements}
One of the co-authors of this work, Prof. David Eichler, sadly passed away during the elaboration of the new version of this manuscript. This paper is dedicated to his memory. 
We would like to thank  with Avishai Gal-Yam, Richard Ellis, Elena Pian and Paolo Mazzali for useful discussions. We acknowledge funding from Israel Science Foundation and John and Robert Arnow Chair of Theoretical Astrophysics.

\end{document}